\documentclass[final]{aipproc}
\usepackage[usenames]{colortbl}
\usepackage{epsfig,rotate}
\usepackage{amsmath}
\usepackage{graphicx}

\layoutstyle{8x11single}

\newcommand{\bc}           {\begin{center}}
\newcommand{\ec}           {\end{center}}

\begin{document}

\title{Baryon Spectroscopy
and the Origin of Mass}


\author{Eberhard Klempt}{
  address={Helmholtz-Institut f\"ur Strahlen-- und Kernphysik,
    Universit\"at Bonn\\
    Nu{\ss}allee 14-16, D-53115 Bonn, GERMANY\\
    \footnotesize\texttt{e-mail: klempt@hiskp.uni-bonn.de}}
}

\begin{abstract}
The proton mass arises from spontaneous breaking of chiral symmetry
and the formation of constituent quarks. Their dynamics cannot be
tested by proton tomography but only by studying excited baryons.
However, the number of excited baryons is much smaller than expected
within quark models; even worse, the existence of many known states
has been challenged in a recent analysis which includes - compared
to older analyses - high-precision data from meson factories. Hence
$\pi N$ elastic scattering data do not provide a well-founded
starting point of any phenomenological analysis of the baryon
excitation spectrum. Photoproduction experiments now start to fill
in this hole. Often, they confirm the old findings and even suggest
a few new states. These results encourage attempts to compare the
pattern of observed baryon resonances with predictions from quark
models, from models generating baryons dynamically from
meson-nucleon scattering amplitudes, from models based on
gravitational theories, and with the conjecture that chiral symmetry
may be restored at high excitation energies. Best agreement is found
with a simple mass formula derived within AdS/QCD. Consequences for
our understanding of QCD are discussed as well as experiments which
may help to decide on the validity of models.
\end{abstract}
\maketitle
 %
\section{Introduction}
Baryon spectroscopy is at a bifurcation point. The Particle Data
Group [1] lists 44 $N$ and $\Delta$ resonances which stem mostly
from the old analyses of the Karlsruhe-Helsinki (KH) [2] and
Carnegie-Mellon (CM) [3]collaborations. The most recent analysis of
the SAID group at George Washington University (GWU) [4] includes
high precision data from meson factories and data with measurements
of the proton recoil polarization. So we should expect the number of
known states to increase. Instead, SAID finds 20 only. This is a
much smaller number than the 72 known mesonic states, not to speak
about the wealth of additional states revealed from the
QMC-St-Petersburg analysis of Crystal-Barrel LEAR data. Of course, a
much greater number of states is expected for (three-body) baryons
than for (two-body) mesons. Here, we will address the question
wether we have to abandon a large fraction of baryon resonances
listed by the PDG or if we should include them when interpreting the
data in models.

The most natural and most popular frame to discuss baryon resonances
is the quark model. But other concepts have been proposed. Here,
data will be compared to quark-model predictions, to the Skyrme
model, to AdS/QCD. At the end, I will address a few key issues in
theory and will propose key experiments, which may help to decide on
the validity of the different concepts.
\section{Why baryon spectroscopy?}
We all have an intuitive knowledge of what ``mass" means. But, as
often in physics, Nature offers surprises. Astronomers believe that
only a small fraction of the total mass of the Universe is matter in
the form as we know it. Close galaxies seem to drift apart faster
than more remote galaxies (Fig. 1, left); the acceleration is
assigned to a mythic dark energy (accounting for 73\% of the mass of
the Universe). The rotational frequencies of stars in galaxies do
not depend on the distance from the galaxy center (Fig. 1, center);
dark matter, e.g. in the form of super-symmetric particles, is
introduced which gives a 23\% mass contribution. ``Normal" matter
makes up the remaining 4\%. The myriads of stars in the nightly
heaven constitute just 1\% of the total mass (Fig. 1, right).

The matter as we know it provides surprises for us as well. We are
accustomed to the fact that electrons carry only an atomic mass
fraction of 1/4000; their binding energy is in the sub-ppm range. In
nucleons, it is the field energy which outcasts the quark mass by a
\begin{figure}[pt]
 \begin{tabular}{ccc}
 \includegraphics[width=4.6cm,height=5.6cm,angle=-90]{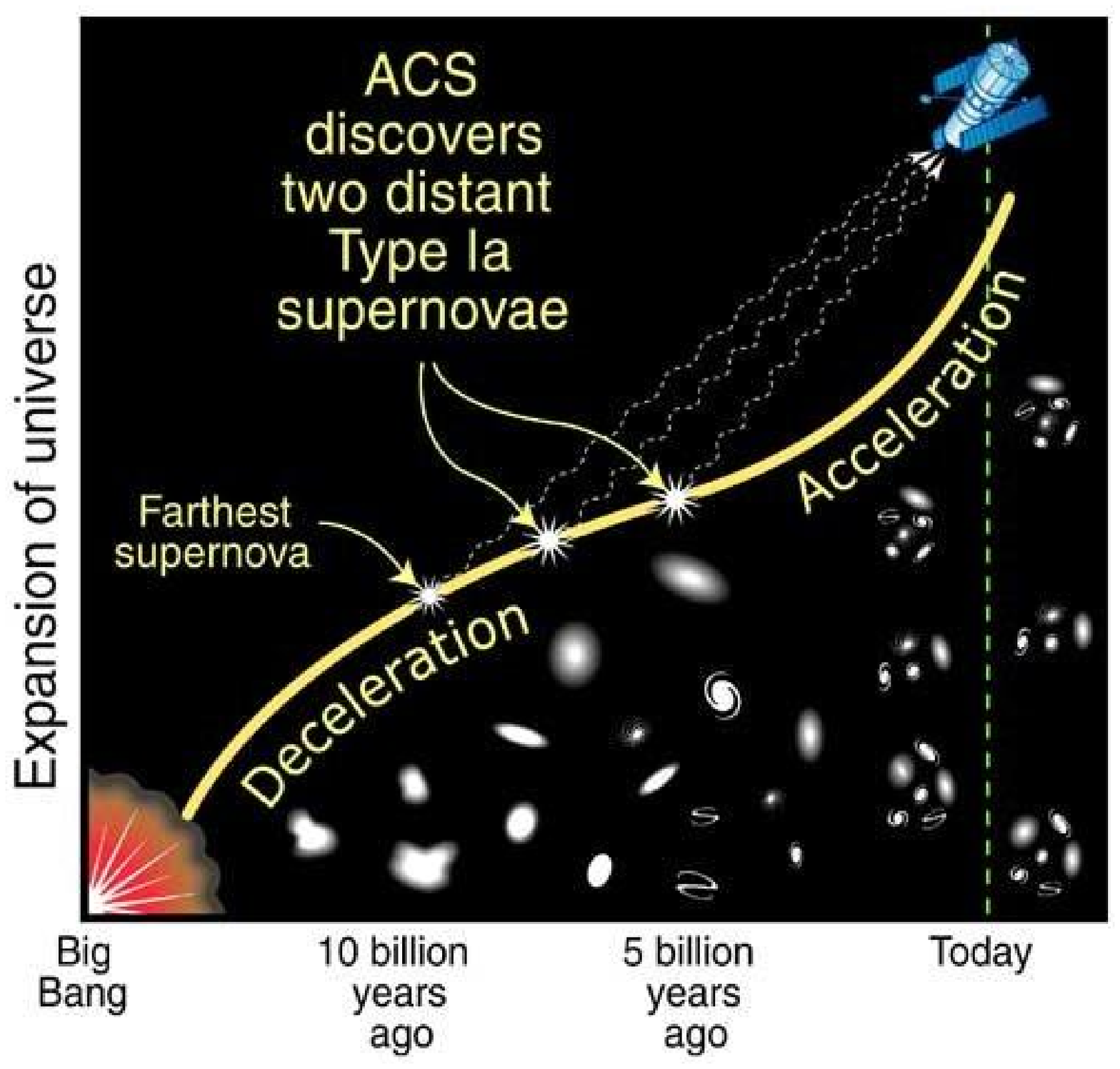}&\hspace{-5mm}
 \includegraphics[width=4.0cm,height=5.4cm,angle=-90]{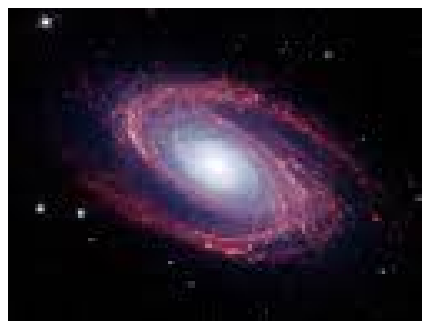}&\hspace{-5mm}
 \includegraphics[width=4.0cm,height=4.8cm,angle=-90]{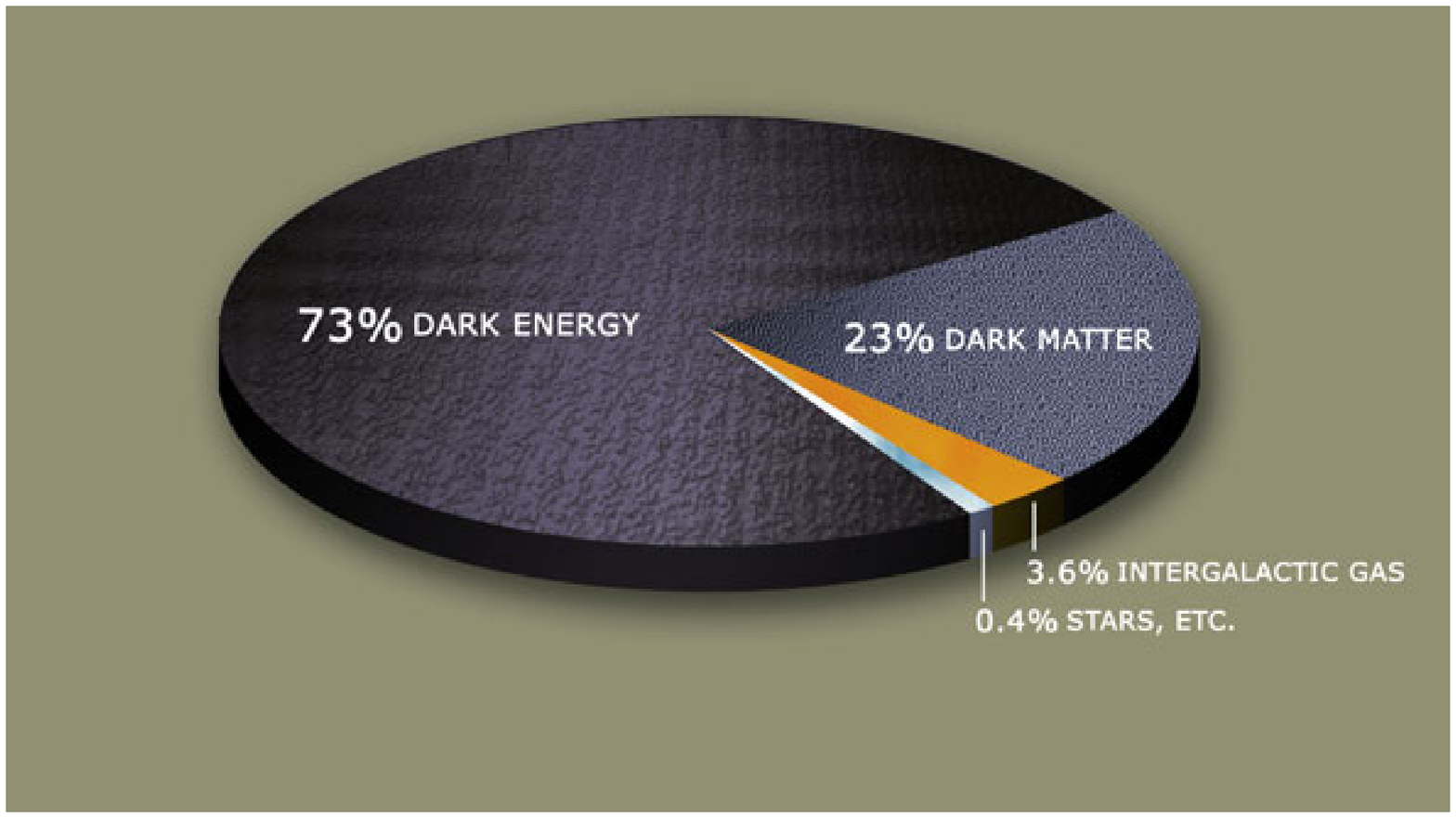}\\[-0.5ex]
  \caption{\it Left: The Hubble constant is larger for close-by galaxies.
  Center: The velocities of stars within galaxies does not de\-pend on
their distance to the center. Right: Mass distribution.  }\\[-3.5ex]
  \end{tabular}
\end{figure}
factor hundred. 99\% of the baryon mass does not come of the quark
masses but from chiral symmetry breaking. The small current quark
mass due to the quark-Higgs interaction is not very important on the
hadronic scale.\\[-3.8ex]

\hspace{-3.5mm}\begin{minipage}[c]{0.60\textwidth}

The fundamental theory of strong interactions, QCD, is chirally
invariant, it keeps handiness, for the nearly massless quarks. This
symmetry is dynamically broken, giving to quarks an effective mass.
This effect can be studied in lattice gauge calculations [5] which
show an increase in mass with decreasing momentum transfer (see Fig.
2). At small $q$, the effective mass is about 320\,MeV. This is
called the constituent quark mass. It can be understood using
different languages. The bag model assigns the constituent quark
mass to absence of quark condensates inside of the bag [6]; the
Dyson Schwinger equation approximates the effective gluon operator
[7]; quarks hop between special field configurations called
instantons by flipping the quark spin and thus, these acquire mass
[8]. How can we learn about constituent quarks? Certainly not by
deep inelastic scattering. Proton
\end{minipage}\hspace{3mm}
\begin{minipage}[c]{0.39\textwidth}
\includegraphics[height=60mm,angle=90]{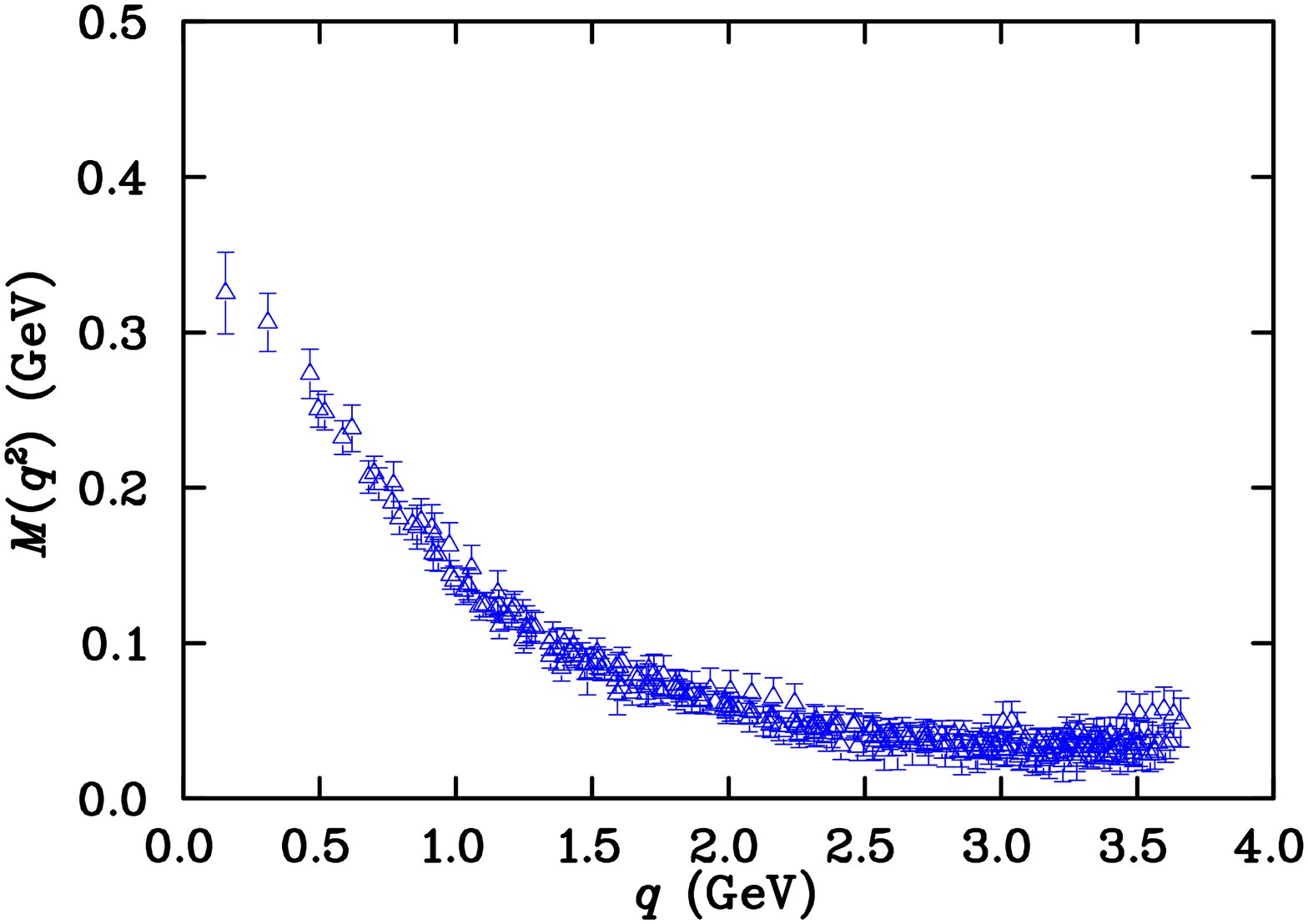}\\
\label{fig:astigm}
{\bf Figure~2.}\it ~The quark mass as a function\\
\phantom{ttttttttttttttt}of the momentum transfer [5]\vspace{8mm}
\end{minipage}\\[-1.9ex]
tomography reveals the distribution of linear and orbital angular
 momenta but information on collective degrees of freedom is
lost. The wave length with which the proton is explored needs to
match the size of the constituents. This is the realm of
spectroscopy. \\[3.0ex]
\hspace{-3.5mm}\begin{minipage}[c]{0.36\textwidth}
\section{Experimental status } The Particle Data
Group~[1] lists 44 nucleon and $\Delta$ resonances, Table 1 presents
a recent compilation [9]. Evi\-dence for the existence of baryon
resonances is derived mostly from elastic $\pi N$ scattering.
Depending on the confidence with which their existence and their
properties are known, these resonances are decorated with one, two,
three or four stars. Except for the four-star resonances, the
evidence is challenged by a careful GWU analysis [4] and only those
boldfaced in Table 1 survive. Spin and parity of resonances are
given in the form $N_{1/2^-}(1535)$ which gives spin and parity
directly instead of $N(1535)S_{11}$ used by the Particle Data Group.
\end{minipage}
\hspace{2mm}\begin{minipage}[c]{0.63\textwidth}\vspace{4mm}
\label{table:all}{\bf Table 1.}{\it N and $\Delta$ resonances.
Compilation: see [9]. $^{a}$: BnGa; $^{b}$:  GWU.}
\\[0.4ex]
\begin{footnotesize} \renewcommand{\arraystretch}{1.24}
\begin{tabular}{cccccc}
\hline
Resonance\hspace{-3mm}&\hspace{-3mm}Mass\hspace{-3mm}&\hspace{-3mm}
Resonance\hspace{-3mm}&\hspace{-3mm}Mass\hspace{-3mm}&\hspace{-3mm}
 Resonance\hspace{-3mm}&\hspace{-3mm}Mass\\
\hline $ N\bf (940)$
\hspace{-3mm}&\hspace{-3mm}940\hspace{-3mm}&\hspace{-3mm}
 $\Delta\bf(1232)$\hspace{-3mm}&\hspace{-3mm}1232\;$\pm$\;1\hspace{-3mm}
 &\hspace{-3mm}$ N{\bf_{1/2^+}(1440)}$\hspace{-3mm}&\hspace{-3mm}1450$\pm$32\\

$
N{\bf_{1/2^-}(1535)}$\hspace{-3mm}&\hspace{-3mm}1538$\pm$10\hspace{-3mm}&\hspace{-3mm}
$
N{\bf_{3/2^-}(1520)}$\hspace{-3mm}&\hspace{-3mm}1522\,$\pm$\,4\hspace{-3mm}&\hspace{-3mm}
$ N{\bf_{1/2^-}(1650)}$\hspace{-3mm}&\hspace{-3mm}1660$\pm$18\\

$
N_{3/2^-}(1700)$\hspace{-3mm}&\hspace{-3mm}1725$\pm$50\hspace{-3mm}&\hspace{-3mm}
$
N{\bf_{5/2^-}(1675)}$\hspace{-3mm}&\hspace{-3mm}1675\,$\pm$\,5\hspace{-3mm}&\hspace{-3mm}
$\Delta\bf_{1/2^-}(1620)$\hspace{-3mm}&\hspace{-3mm}1626$\pm$23\\

$\Delta\bf{
_{3/2^-}(1700)}$\hspace{-3mm}&\hspace{-3mm}1720$\pm$50\hspace{-3mm}&\hspace{-3mm}
$\Delta
_{3/2^+}(1600)$\hspace{-3mm}&\hspace{-3mm}1615$\pm$80\hspace{-3mm}&\hspace{-3mm}
$ N{\bf_{3/2^+}(1720)}$ \hspace{-3mm}&\hspace{-3mm}1730$\pm$30\\

$N{\bf_{5/2^+}(1680)}$
\hspace{-3mm}&\hspace{-3mm}1683\,$\pm$\,3\hspace{-3mm}&\hspace{-3mm}
$ N_{1/2^+}(1710)$
\hspace{-3mm}&\hspace{-3mm}1713$\pm$12\hspace{-3mm}&\hspace{-3mm}
$\Delta_{1/2^+}(1750)$\hspace{-3mm}&\hspace{-3mm}\\

$
N_{1/2^-}(1905)$\hspace{-3mm}&\hspace{-3mm}1905$\pm$50\hspace{-3mm}&\hspace{-3mm}
$
N_{3/2^-}(1860)$\hspace{-3mm}&\hspace{-3mm}1850$\pm$40\hspace{-3mm}&\hspace{-3mm}
\hspace{1.5mm}$ N_{1/2^+}(1880)^a$\\

$
N_{3/2^+}(1900)^a$\hspace{-3mm}&\hspace{-3mm}\hspace{-3mm}&\hspace{-3mm}
\hspace{1.5mm}$
N_{5/2^+}(1910)$\hspace{-3mm}&\hspace{-3mm}1880$\pm$40\hspace{-3mm}&\hspace{-3mm}
$ N_{7/2^+}(1990)$\hspace{-3mm}&\hspace{-3mm}2020$\pm$60\\

$\Delta
_{1/2^-}(1900)$\hspace{-3mm}&\hspace{-3mm}1910$\pm$50\hspace{-3mm}&\hspace{-3mm}
$\Delta
_{3/2^-}(1940)$\hspace{-3mm}&\hspace{-3mm}1995$\pm$60\hspace{-3mm}&\hspace{-3mm}
$\Delta\bf{
_{5/2^-}(1930)}$\hspace{-3mm}&\hspace{-3mm}1930$\pm$30\\

$\Delta\bf{
_{1/2^+}(1910)}$\hspace{-3mm}&\hspace{-3mm}1935$\pm$90\hspace{-3mm}&\hspace{-3mm}
$\Delta{
_{3/2^+}(1920)}$\hspace{-3mm}&\hspace{-3mm}1950$\pm$70\hspace{-3mm}&\hspace{-3mm}
$\Delta\bf{
_{5/2^+}(1905)}$\hspace{-3mm}&\hspace{-3mm}1885$\pm$25\\

$\Delta\bf{
_{7/2^+}(1950)}$\hspace{-3mm}&\hspace{-3mm}1930$\pm$16\hspace{-3mm}&\hspace{-3mm}
$ N_{1/2^+}(2100)$\hspace{-3mm}&\hspace{-3mm}2090$\pm$100\hspace{-3mm}&\hspace{-3mm} $ N_{1/2^-}(2090)$\hspace{-3mm}&\hspace{-3mm}\\

$
N_{3/2^-}(2080)$\hspace{-3mm}&\hspace{-3mm}2100$\pm$55\hspace{-3mm}&\hspace{-3mm}
$
N_{5/2^-}(2060)^{a}$\hspace{-3mm}&\hspace{-3mm}2065$\pm$25\hspace{-3mm}&\hspace{-3mm}
$\bf
N_{7/2^-}(2190)$\hspace{-3mm}&\hspace{-3mm}2150$\pm$30\\

$
N_{5/2^-}(2200)$\hspace{-3mm}&\hspace{-3mm}2160$\pm$85\hspace{-3mm}&\hspace{-3mm}
$\bf
N_{9/2^-}(2250)$\hspace{-3mm}&\hspace{-3mm}2255$\pm$55\hspace{-3mm}&\hspace{-3mm} $\Delta _{1/2^-}(2150)$\hspace{-3mm}&\hspace{-3mm}\\

\hspace{1.5mm}$\Delta\bf
_{5/2^-}(2223)^b$\hspace{-3mm}&\hspace{-3mm}\hspace{-3mm}&\hspace{-3mm}
$\Delta
_{7/2^-}(2200)$\hspace{-3mm}&\hspace{-3mm}2230\,$\pm$\,50\hspace{-3mm}&\hspace{-3mm}
$
N{\bf_{9/2^+}(2220)}$\hspace{-3mm}&\hspace{-3mm}2360$\pm$125\\

$\Delta
_{7/2^+}(2390)$\hspace{-3mm}&\hspace{-3mm}2390$\pm$100\hspace{-3mm}&\hspace{-3mm}
$\Delta
_{9/2^+}(2300)$\hspace{-3mm}&\hspace{-3mm}2360$\pm$125\hspace{-3mm}&\hspace{-3mm}
\hspace{1mm}$\Delta\bf{
_{11/2^+}(2420)}$\hspace{-3mm}&\hspace{-3mm}2462$\pm$120\\

$\boldmath\Delta
_{9/2^-}(2400)$\hspace{-3mm}&\hspace{-3mm}2400$\pm$190
\hspace{-3mm}&\hspace{-3mm} $\Delta
_{3/2^-}(2350)$\hspace{-3mm}&\hspace{-3mm}2310\,$\pm$\,85\hspace{-3mm}&\hspace{-3mm}
$
N_{11/2^-}(2600)$\hspace{-3mm}&\hspace{-3mm}2630$\pm$120\\

$
N_{13/2^+}(2800)$\hspace{-3mm}&\hspace{-3mm}2800$\pm$160\hspace{-3mm}&\hspace{-3mm}
\hspace{1mm}$\Delta _{13/2^-}(2750)$\hspace{-3mm}&\hspace{-3mm}
2720$\pm$100\hspace{-3mm}&\hspace{-3mm}\hspace{1mm}$\Delta
_{15/2^+}(2950)$
\hspace{-3mm}&\hspace{-3mm}2920$\pm$100\\
\hline\\[-1.8ex]
\end{tabular}
\end{footnotesize}\hspace{-3mm}\renewcommand{\arraystretch}{1.0}
\end{minipage}
For an interpretation of the resonance spectrum it is decisive to
know if we need to abandon the 24 states not seen by SAID. At
present, we cannot decide if the KH and CM analyses pick up some
noise or if resonances are lost in the GWU analysis by the applied
smoothing procedure. But for a few resonances, seen in the KH and CM
analyses and not seen in the GWU analysis of elastic $\pi N$
scattering data, this question can be studied. We give three
examples where resonances found by KH and CM and disclaimed by GWU
are required in inelastic reactions. Note that the relevant $\pi
N\to \pi N$ amplitudes [4] are included in the fit. KH, CM, and GWU
fit only these amplitudes; the inelasticity is unconstrained by
data. In the BnGa analysis, most inelastic channels are known from
photoproduction and only few inelastic channels (mainly with vector
mesons) are treated as ``black box".
\\[0.5ex]

\hspace{-3.5mm}\begin{minipage}[c]{0.65\textwidth} {\bf The
\boldmath$\Delta_{3/2^+}$(1600) from $\pi^+p\to \Sigma^+K^+$:\\ }
Fig. 3 shows the imaginary part of the elastic $P_{33}$ $\pi N$
scattering amplitude above $\Delta(1232)$. Points (in red) with
error bars represent the GWU amplitude, the thin (red) line a BnGa
fit. The $\Delta(1232)$ tail is followed by a continuum; there is no
evidence for further structures. The KH amplitude is represented by
black triangles; the dominant $\Delta(1232)$ tail is removed. The
amplitude exhibits a peak in the imaginary part, evidencing a
$\Delta_{3/2^+}$ resonance at 1600\,MeV coupling to $\pi N$. The CM
partial wave agrees with KH in exhibiting a peak structure. The
question is which analysis is correct the old analyses by CM and KH
finding $\Delta_{3/2^+}$(1600) or GWU where the resonance does not
exist. The amplitudes derived from elastic $\pi N$ scattering are
ambiguous. The BnGa fit includes data on the inelastic reaction
$\pi^+p\to \Sigma^+K^+$. Due to the charge, intermediate resonances
must have isospin $I=3/2$. The data were included in a general fit
to a large number
\end{minipage}
\begin{minipage}[c]{0.34\textwidth}
\includegraphics[width=5.5cm,height=5cm]{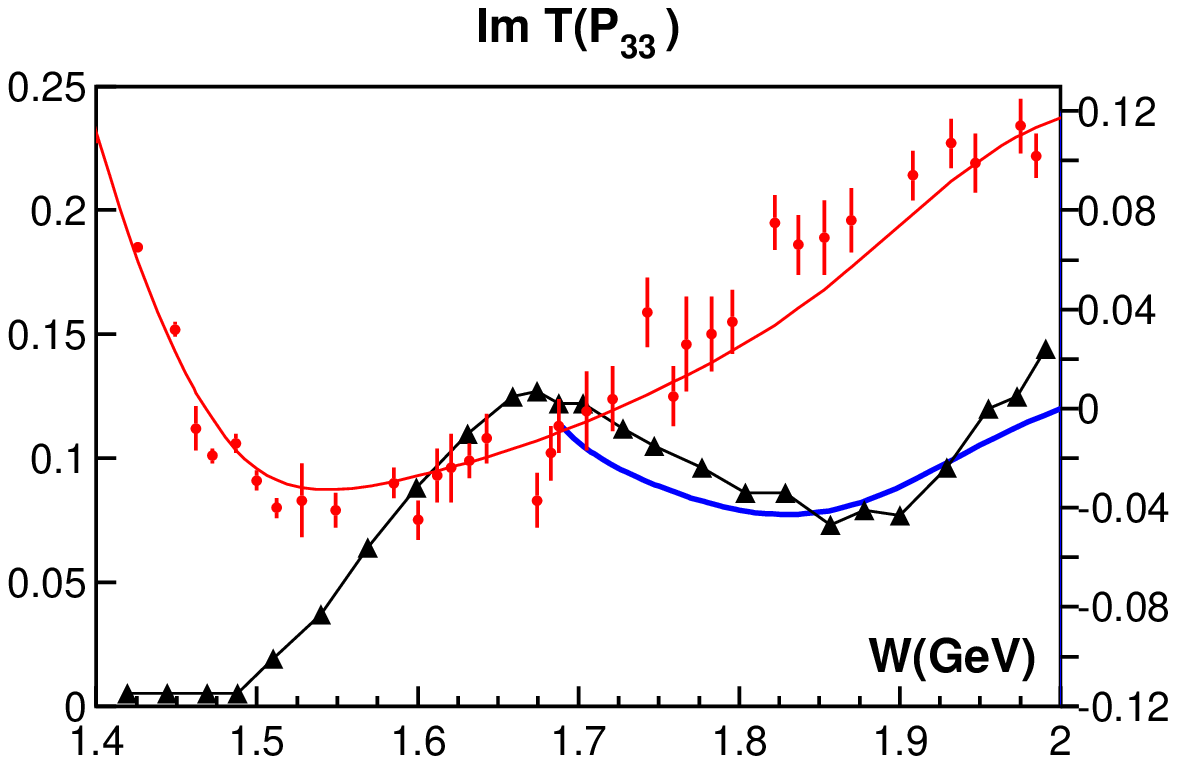}\\[-5ex]
\begin{center}
{\bf Figure~3.}{\it ~The $P_{33}$ amplitude.}\vspace{4mm}
\end{center}
\end{minipage}
of different inelastic reactions and to the GWU elastic amplitudes.
The resulting $P_{33}$ amplitude is shown as thick (blue) curve. The
amplitude follows closely the KH amplitude.  The analysis requires
to introduce $\Delta_{3/2^+}(1600)$. The need for the resonance can
be seen from the differential distribution and the induced
$\Sigma^+$ polarization. Thus we believe that the
$\Delta_{3/2^+}(1600)$ is definitely confirmed with a pole position
at $M=1540^{+40}_{-80}$, $ \Gamma=230\pm40$\,MeV. \\[1.5ex]
\hspace{-3.5mm}\begin{minipage}[c]{0.32\textwidth}
\includegraphics[width=5cm]{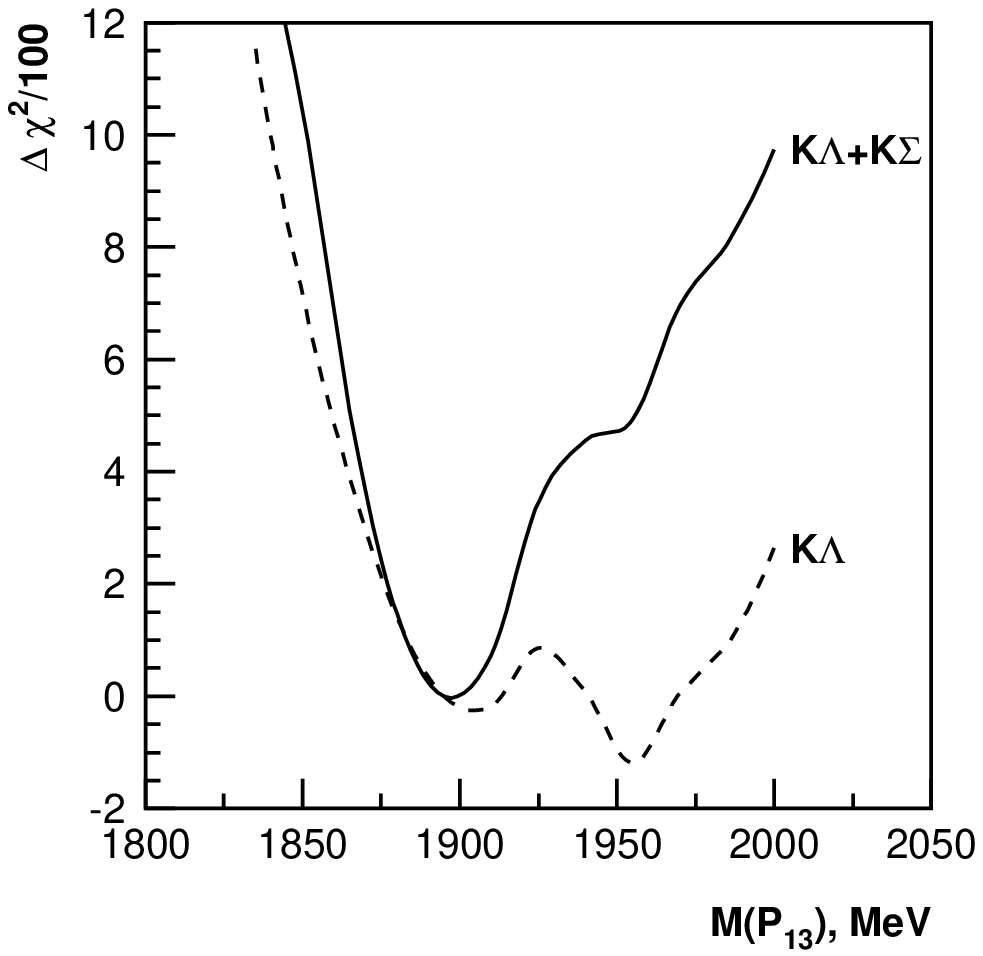}\\
{\bf Figure~4.}{\it~$\chi^2$ as a function of the\\ \phantom{zzzzz}
imposed $ N_{3/2^+}(1900)$ mass}.
\end{minipage}
\begin{minipage}[c]{0.68\textwidth}
{\bf The \boldmath$ N_{3/2^+}$(1900) from photoproduction of
hyperons:}\\ Very sensitive data on photoproduction of hyperons are
now available on $\gamma p\to \Lambda K^+$, $\gamma p\to \Sigma^0
K^+$, and $\gamma p\to \Sigma^+ K^0$. Fig. 4 shows the $\chi^2$ of
the BnGa fit as a function of the assumed $N_{3/2^+}(1900)$ mass.
The data base includes high-statistics angular distributions,
several single ($\Sigma, T$ and $ P$) and double polarization
observables ($ C_x, C_z, O_x, O_z$). A large data base on other
reactions is included in the fits. But still, there is not yet a
full reconstruction of partial wave amplitude. So the evidence is
derived from a $\chi^2$ minimization (see Fig. 4). The best values
and errors, $M=\,1915\pm 50$\,MeV and $\Gamma = 100\pm50$\,MeV,
cover all solutions with reasonable $\chi^2$.\\[1.5ex]
{\bf \boldmath$\Delta_{3/2^+}$(1920) and $\Delta_{3/2^-}$(1940) from
$\gamma p\to p\pi^0\eta$:\unboldmath\\ }
Fits to this reaction with/without $\Delta_{3/2^+}(1920)$ (left) or
$\Delta_{3/2^-}(1940)$ (right) represented by solid lines in Fig. 5
are not not satisfying, both these resonances are needed to achieve
a good fit. The fits optimizes for $(M; \Gamma) =($1910$\pm$50;
330$\pm50)$ ($\Delta_{3/2^+}(1920)$) and (1985$\pm$30;
390$\pm$50)\,MeV ($\Delta_{3/2^-}(1940)$).\\[-1ex]
\end{minipage}

\hspace{-5mm}\begin{tabular}{cc}
\includegraphics[width=8.cm]{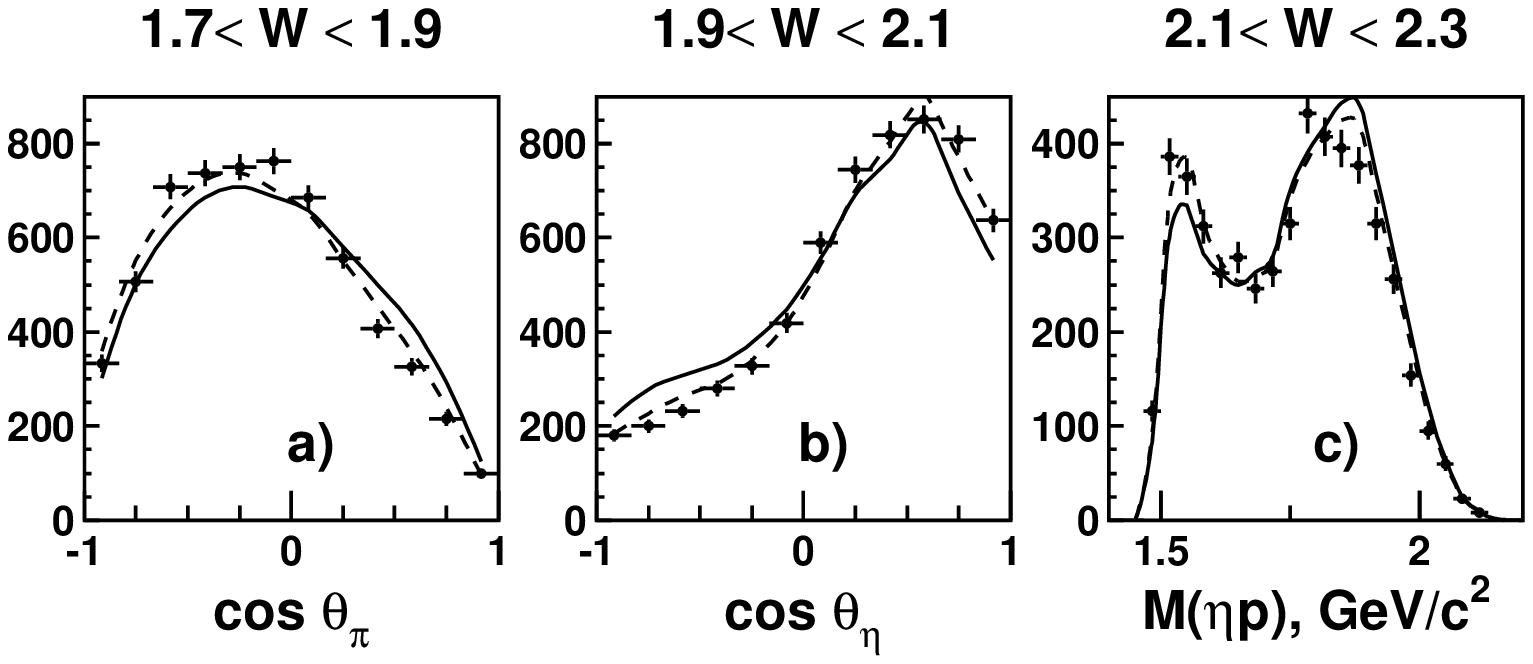}&
\hspace{-5mm}\includegraphics[width=8.cm]{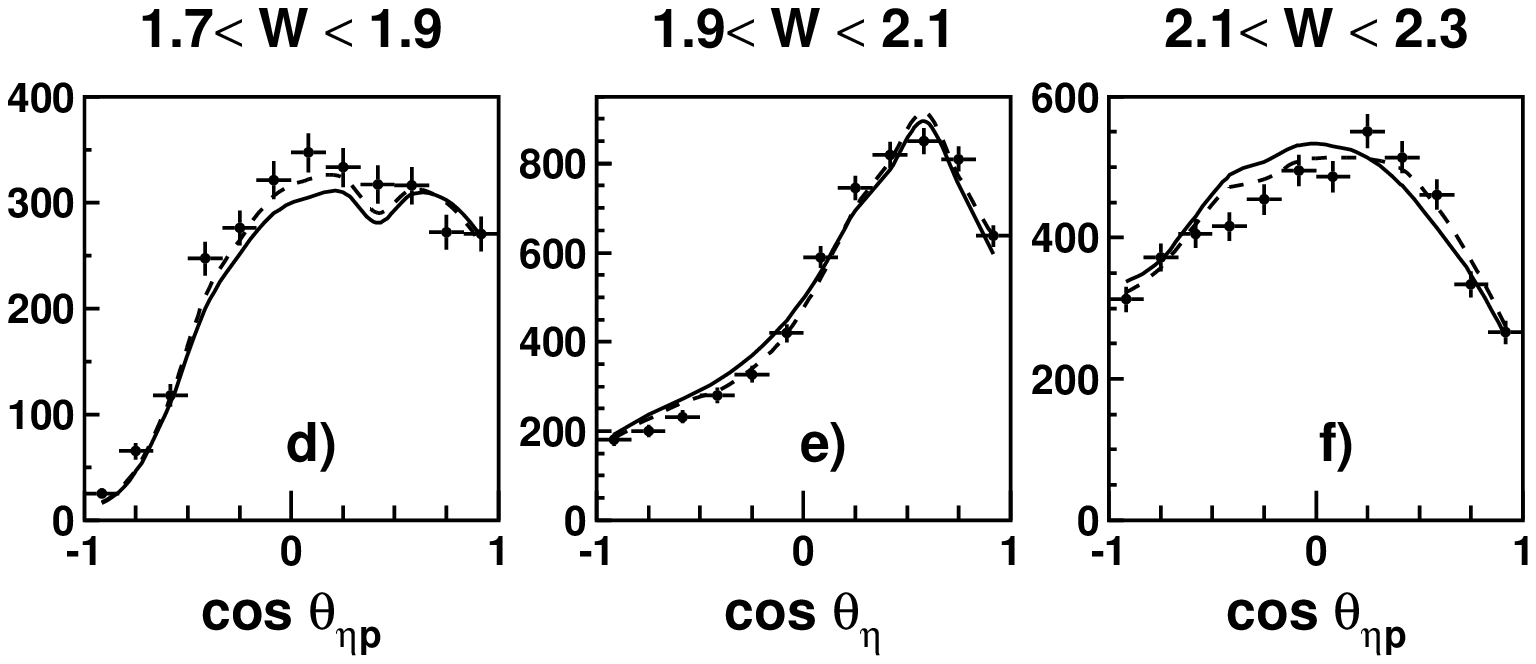}
\end{tabular}
{\bf Figure 5:}{\it Selected mass and angular distributions. A fit
represented by solid lines without $\Delta_{3/2^+}(1920)$ (left) or
$\Delta_{3/2^-}(1940)$ (right) yields a bad fit. Only when both
resonances are included (dashed lines), data are reproduced.}

We conclude that {\bf 1.} it would be too early to abandon the
resonances found in the KH and CM analyses but missed in the GWU
analysis, and that {\bf 2.} photoproduction begins to make a
significant impact on baryon spectroscopy. For the interpretation of
the baryon spectrum, all resonances of Table 1 will be used. For
further information on the BnGa analysis, see talks by Anisovich and
Sarantsev at this conference.\\[-2ex]

\section{Interpretation}
{\bf Quark models: \\[0.5ex]} The quark model provides the most natural and
most accepted picture of the baryon excitation spectrum. Ingredients
are constituent quarks with defined rest masses, a confinement
potential (mostly linear)  and some residual interaction. In the
celebrated Isgur-Karl model and its later ``relativized"
refinements, an effective one-gluon exchange is chosen as residual
interaction. The Bonn group starts from a Bethe-Salpeter equation;
the linear confinement potential has a full Dirac structure.
Instanton-induced interactions are responsible for the $N - \Delta$
mass splitting. The quark models are very successful in explaining
the properties of baryon ground states and low-mass excitations;
they fail to reproduce the masses of radial excitations
and predict many more baryon resonances than found in experiment. \\[1.ex]
\noindent{\bf AdS/QCD:\\[0.5ex]} The Maldacena correspondence relates
conformal strongly-coupled theories in space-time to a
weakly-coupled (``gravitational") theory in a five dimensional
Anti-de Sitter space embedded in six dimensions. Gravitational
theories can be solved analytically, the solutions can be mapped
into space-time and compared to data. There is a (heuristic) mapping
of quantum mechanical operators to operators in Ads/QCD. The fifth
dimension in AdS called $z$ can be interpreted as virtuality or
distance between constituents. For $z\to 0$, constituents are
asymptotically free. Confinement can be enforced by a hard boundary
$z<z_{\rm max}=1/\Lambda_{\rm QCD}$ (hard wall) or by a soft wall
due to a dilation field (or penalty function) increasing as $z^2$.

AdS/QCD relates masses to the orbital angular momentum $L$ (Fig. 6).
In quark models, relativity plays an important role, and only the
total angular momentum $J$ is defined. Experimentally, there are a
few striking examples where the leading orbital angular momentum and
the spin can be identified. To give one example: the four states
$\Delta_{1/2^+}(1910)$, $\Delta_{3/2^+}(1920)$, $\Delta
_{5/2^+}(1905)$, and $\Delta _{7/2^+}(1950)$ are isolated in mass.
They obviously form a spin-quartet of resonances with $L=2, S=3/2$.
Small admixtures of other components are not excluded.

\hspace{-3mm}\begin{minipage}[c]{0.37\textwidth}

{\bf Figure 6:} {\it One-parameter fit to the $\Delta$ excitation
spectrum. For nucleon resonances, a term is needed which reduces the
mass of baryons with ``good diquarks", with a pair of quarks with
spin and isospin zero. \\[-7.5ex]

\bc\begin{eqnarray}\bf M^2 = a\cdot ({L}+{N}+3/2)
\bf -\,b\cdot \alpha_D\,\left[{\mathrm
GeV^2}\right]\nonumber\\[-3.5ex]\nonumber
\end{eqnarray}
$a=1.04$\,GeV$ ^2$ and $ b=1.46$\,GeV$ ^2$.\ec $\Delta$ resonances
have no good diquarks.}
\end{minipage}
\begin{minipage}[c]{0.62\textwidth}
\includegraphics[width=10.5cm,height=4.8cm]{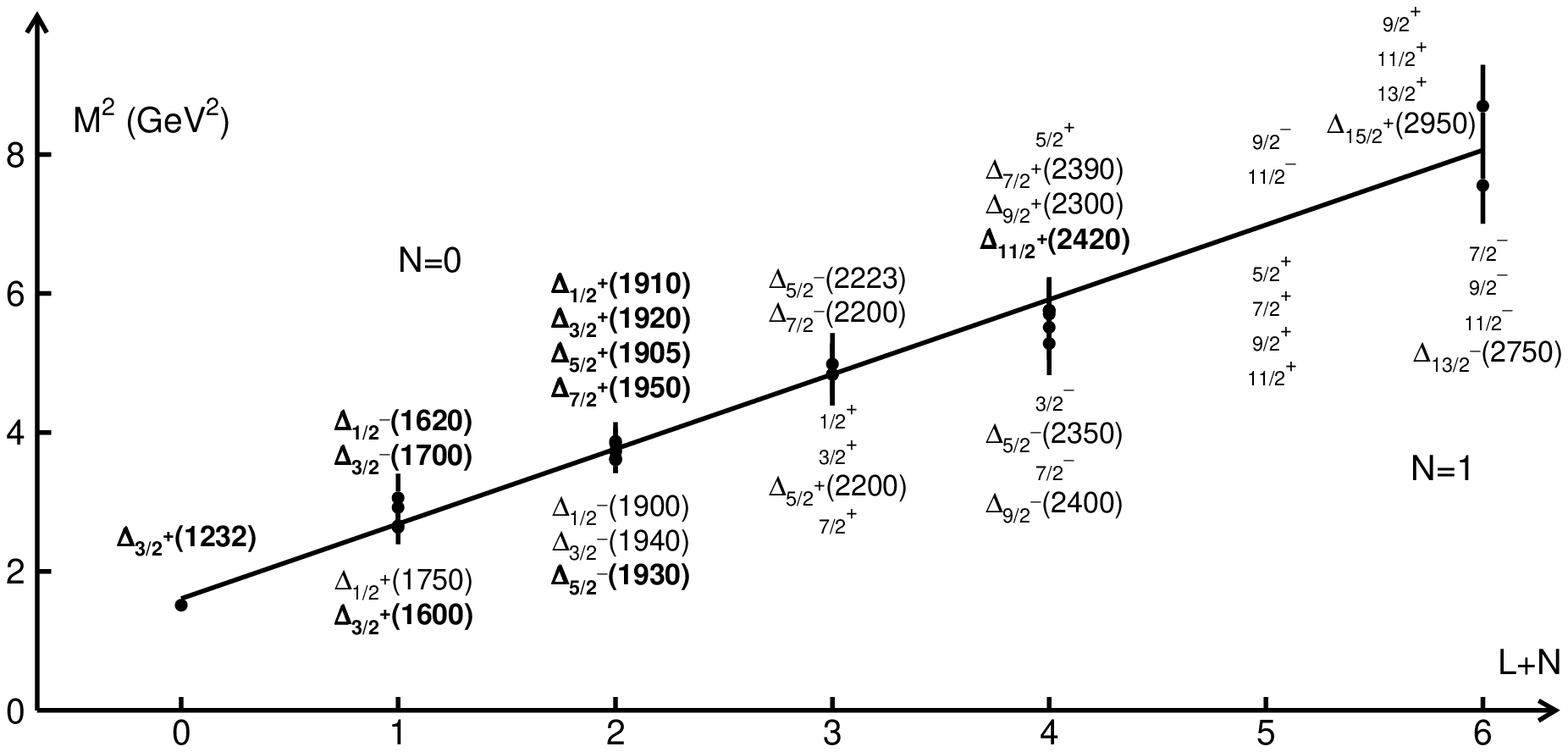}
\end{minipage}
\noindent{\bf Dynamically generated resonances: \\[0.5ex]} The properties of
some resonances (and the meson-baryon system at low energies) can be
understood very successfully using an effective chiral Lagrangian,
relying on an expansion in increasing powers of meson masses and
momenta. A problem arises when these dynamically generated
resonances are predicted atop the quark-model states, or when the
light quark baryons are disregarded altogether and replaced by a
systematics of meson--baryon excitations. The pole structure of
$N_{1/2^-}(1535)$ and $N_{1/2^-}(1650)$, e.g., was studied by
D\"oring {\it et al.} [10]. They generated $N_{1/2^-}(1535)$
dynamically and introduced two additional poles, one for
$N_{1/2^-}(1650)$ and a third one. The latter pole moves far into
the complex plane and provides an almost energy independent
background while the dynamically generated $N_{1/2^-}(1535)$ pole
appears as a stable object. This could mean that $N_{1/2^-}(1535)$
is fully understood by the interaction of the baryon and meson, into
which it disintegrates. It can be interpreted that the observed
$N_{1/2^-}(1535)$ is dynamically generated and the state predicted
by the quark model is missing. The latter interpretation seems
unacceptable to me.\\[1.5ex]
\noindent{\bf Restoration of chiral symmetry?\\[0.5ex]}
In contrary to expectations based on the quark model in harmonic
oscillator approximation, the masses of baryon resonances do not
increase with alternating states of even and odd parity; often,
states having the same $J$ but opposite parities are approximately
degenerate in mass (see Table 2). At low mass, $N_{1/2^-}(1535)$ is
much heavier than its chiral partner, the proton. In meson
spectroscopy, the $\rho$ mass is much below the $a_1(1260)$ mass.
The mass difference is assigned to a spontaneous breaking of chiral
symmetry. Glozman [11] and others argue that at high-mass
excitations, details of the potential responsible for spontaneous
chiral breaking are irrelevant: chiral symmetry could be restored in
highly excited baryons. The alternative interpretation of the
occurrence spin-parity\\[0.5ex]
\hspace{-3.5mm}\begin{minipage}[c]{0.41\textwidth} partners,
AdS/QCD, is criticized because of the use of orbital angular
momentum, a non-relativistic quantity [12], a view contested in
[13]. Parity doublets are only observed for resonances on Regge
daughter trajectories; mesons and baryons falling on the leading
Regge trajectory have no parity partner. This
is a challenge for future experiments.\\[2ex]
\noindent{\bf Models versus data:\\[0.5ex]}
Model predictions can be confronted with data. The predictions of
chiral symmetry restoration give an interpretation of one
observation, that baryons often appear as parity doublets. There is
no prediction where the masses should be found. For this reason, we
do not include this
\end{minipage}\hspace{3mm}
\begin{minipage}[c]{0.55\textwidth}
\renewcommand{\arraystretch}{1.3}
{\bf Table 2.} {\it Chiral multiplets (doublets or quartets) of
$N^*$ and $\Delta^*$ resonances of high mass. List and star rating,
see [9].}\\[0.5ex]
\renewcommand{\arraystretch}{1.2}
\begin{footnotesize}
\begin{tabular}{ccccc} \hline\hline\\[-3ex]
$J$=$\frac{1}{2}$&N$_{1/2^+}(1710)$&\hspace{-0.2mm}N$_{1/2^-}(1650)$&\hspace{-0.2mm}$\Delta_{1/2^+}(1750)$&\hspace{-0.2mm}$\Delta_{1/2^-}(1620)$\vspace{-1.5mm}\\
&***&****&&****\vspace{-1mm}\\
$J$=$\frac{3}{2}$&N$_{3/2^+}(1720)$&\hspace{-0.2mm}N$_{3/2^-}(1700)$&\hspace{-0.2mm}$\Delta_{3/2^+}(1600)$&\hspace{-0.2mm}$\Delta_{3/2^-}(1700)$\vspace{-1.5mm}\\
&****&***&***&****\vspace{-1mm}\\
$J$=$\frac{5}{2}$&N$_{5/2^+}(1680)$&\hspace{-0.2mm}N$_{5/2^-}(1675)$&\multicolumn{2}{c}{no chiral partners}\vspace{-1.5mm}\\
&****&****&&\vspace{-1mm}\\
$J$=$\frac{1}{2}$&N$_{1/2^+}(1880)$&\hspace{-0.2mm}N$_{1/2^-}(1905)$&\hspace{-0.2mm}$\Delta_{1/2^+}(1910)$&\hspace{-0.2mm}$\Delta_{1/2^-}(1900)$\vspace{-1.5mm}\\
&**&*&****&**\vspace{-1mm}\\
$J$=$\frac{3}{2}$&N$_{3/2^+}(1900)$&\hspace{-0.2mm}N$_{3/2^-}(1860)$&\hspace{-0.2mm}$\Delta_{3/2^+}(1920)$&\hspace{-0.2mm}$\Delta_{3/2^-}(1940)$\vspace{-1.5mm}\\
&*&**&***&**\vspace{-1mm}\\
$J$=$\frac{5}{2}$&N$_{5/2^+}(1870)$&\hspace{-0.2mm}no ch. partner\hspace{-0.2mm}&\hspace{-0.2mm}$\Delta_{5/2^+}(1905)$&\hspace{-0.2mm}$\Delta_{5/2^-}(1930)$\vspace{-1.5mm}\\
&**&&****&**\vspace{-1mm}\\
$J$=$\frac{7}{2}$&N$_{7/2^+}(1990)$&\hspace{-0.2mm}no ch. partner\hspace{-0.2mm}&\hspace{-0.2mm}$\Delta_{7/2^+}(1950)$&\hspace{-0.2mm}no ch. partner\vspace{-1.5mm}\\
&**&&****&\vspace{-1mm}\\
$J$=$\frac{7}{2}$&\hspace{-0.2mm}no ch. partner\hspace{-0.2mm}&\hspace{-0.2mm}N$_{7/2^-}(2190)$&\hspace{-0.2mm}no ch. partner&\hspace{-0.2mm}$\Delta_{7/2^-}(2200)$\vspace{-1.5mm}\\
&&****&&*\vspace{-1mm}\\
$J$=$\frac{9}{2}$&N$_{9/2^+}(2220)$&\hspace{-0.2mm}N$_{9/2^-}(2250)$&\hspace{-0.2mm}$\Delta_{9/2^+}(2300)$&\hspace{-0.2mm}$\Delta_{9/2^-}(2400)$\vspace{-1.5mm}\\
&****&****&**&**\\
\hline\hline
\end{tabular}
\end{footnotesize} \vspace*{2mm}
\renewcommand{\arraystretch}{1.0}
\end{minipage}\\[-0.4ex]

\noindent model in the quantitative comparison. Also, models
generating resonances dynamically are not suitable for a numerical
comparison with data, since only a part of the spectrum is
calculated. Well suited are quark models, AdS/QCD, and the Skyrme
model (even though this model was left out in the above discussion
of models). We use two quark models for the comparison, the
relativized quark model of Capstick and Isgur [14] and the
relativistic Bonn model [15]. The Skyrme model [16] has two
parameters only but predicts less than half the number of the
observed states, and the mass predictions are rather inaccurate. The
best agreement is achieved with
the~gravitational~model~[17],~see~Table~3.~A~breakdown~of~contributions~of~individual~resonances~can~be~found~in~[18].

The agreement between AdS/QCD and the data is absolutely amazing.
Obviously, the two parameters used to describe the data correspond
to important physical quantities. The parameter $a$ in front of
$L+N$ is related to the maximum distance between constituents, it is
related to the size of the baryon. The second parameter $b$ is used
to construct an operator in AdS which reduces the size proportional
to the diquark fraction with vanishing spin and isospin in the
baryon: in this version of AdS/QCD, ``good diquark" are smaller in
size compared to other diquarks.

The second observation is that the mass depends on the orbital
angular momentum $L$ implying that we have a non-relativistic
situation. We are used to describe the nucleon as bound state of
three constituent quarks each having 1/3 of the nucleon mass. The
constituent-quark is generated by chiral symmetry breaking, by the
energy of the QCD fields. The quark model assumes that the
constituent quark is an object which can be accelerated to
relativistic energies; chiral symmetry and chiral symmetry breaking
is not effected. This does not need to be the case. Glozman assumes
that chiral symmetry is restored. This is the assumption of a
central Mexican-hat-like potential where chiral symmetry is
dynamically broken at the origin of the hadron. The use of a mean
field which leads to a Mexican-hat potential in the rest frame of
the nucleon is at least a debateable assumption. The success of
AdS/QCD suggests that constituent quarks expand and become more
massive when a baryon is excited to high energy. Chiral symmetry is
broken in an extended volume, and this might be the reason for the
increase in mass.\\[1.5ex] {\bf
Table 3.} {\it Comparison of model calculations with the mass
spectrum of nucleon and $\Delta$ resonances from Table 1.} \\[-2.5ex]
\renewcommand{\arraystretch}{1.2}\renewcommand{\arraystretch}{1.2}
\begin{center}
\begin{tabular}{lccc} \hline\hline Model & Reference & Nr. of parameters & ``quality"\\
 Quark model with eff. one-gluon exchange&[14]& 7
 &  $ (\delta M/M) = 5.6$\% \\
Quark model with instanton induced forces  &[15]&
5& $ (\delta M/M) = 5.1$\%\\
Skyrme model:&[16]&2& $ (\delta M/M) = 9.1$\%\\
AdS/QCD model with ``good diquarks": &[17]&2& $(\delta M/M) = 2.5$\%\\
\hline\hline
\end{tabular}\\[1.5ex]
\end{center}
\renewcommand{\arraystretch}{1.0}

\noindent The nucleon is lighter than $\Delta(1232)$, not because of
effective-one gluon exchange leading to a magnetic hyperfine
splitting but because of the smaller size of the good diquark it
contains (which 50\% probability). The $\Lambda(1405)$ has such a
low mass since it has not only one good diquark but all three pairs
have vanishing spin and isospin.
\\

\noindent {\bf Multiplicity of resonances and the existence glueballs and hybrids: \\[0.5ex]}
There is the well-known problem of the {\it missing resonances}: the
number of baryon resonances predicted in quark models exceeds by far
the number of observed states. The number of baryon resonances can
be counted using harmonic oscillator wave function, with two
oscillators $\rho$ and $\lambda$. Quark models predict, for given
$\vec L$ and $N$, a multitude of states satisfying $\vec
l_{\rho}+\vec l_{\lambda}=\vec L$ and $n_{\rho}+n_{\lambda}=N$. With
increasing $L$ and $N$, the number of predicted states explodes:
expected are, e.g. two $N_{1/2^+}$ states in the second shell, seven
$N_{1/2^-}$ states in the third shell, ten $N_{1/2^+}$ states in the
forth shell. This problem is partly cured in AdS/QCD where at most
two $J=1/2$ states are expected in any shell.

There could be a second problem of too large a number of predicted
states. In the quark model, there is e.g. one nucleon resonance with
$L=1$, $S=1/2$, one state with $L=1$, $S=3/2$. The possibility of
mixing admitted, we can still identify the $N_{1/2^-}(1535)$ with
the predicted $L=1$, $S=1/2$ quark model state, and
$N_{1/2^-}(1650)$ with $L=1$, $S=3/2$. But there is evidence that
$N_{1/2^-}(1535)$ can be generated dynamically from $N\eta$-$\Lambda
K$-$\Sigma K$ coupled-channel scattering dynamics. Hence there could
be a quark model state $N_{1/2^-}(1535)$ and a dynamically generated
$N'_{1/2^-}(1535)$. Zou [18] proposes that the observed
$N_{1/2^-}(1535)$ has a large $(qqq)(q\bar q)$ component with all
quarks in S-wave. If $N_{1/2^-}(1535)= \alpha|qqq> +
\beta|(qqq)(q\bar q)>$, is there an orthogonal resonance
$N''_{1/2^-}= \beta|qqq> - \alpha^*|(qqq)(q\bar q)>$ and, if so, at
which mass? Do hybrid configurations $N'''_{1/2^-}=\alpha'|qqq> +
\beta'|(qqq)(G)>$ add to the list of expected resonances?
Experimentally, none of these additional states has been observed.

It seems to be worthwhile to stress that the number of bosons is not
a well-defined quantity. I propose a view in which the interactions
between three (current) quarks provide the primary forces to
stabilize a baryon. The three quarks acquire their constituent
dynamical mass. Gluons may polarize the vacuum, correlated
quark-antiquark pairs are created. Depending on the dynamics, these
$q\bar q$ pairs may have long-range correlations and may move freely
within a hadron as Cooper-pairs or massless Goldstone bosons leading
to a fast flavor exchange. Or they may evolve into massive quarks.
If their masses approach the mass of ``normal" constituent quarks,
we rather speak about a nucleon-meson molecule or - invoking color
chemistry - about five-quark states. However, the origin of all
baryons is the three-quark component. The actual decomposition is a
question how the QCD vacuum responds to the primary color source of
three quarks at a given mass and with given quantum numbers. The
response can be very different, in particular it will depend
critically on the presence of close-by thresholds, but it is a
unique response. There is no ``hidden variable" which decides that
three quarks with $L=1$, $S=1/2$ will go either into a $N$, $N'$,
$N''$, or into a $N{'''}$ configuration, or into different mixtures
of these Fock components. There is only one state.

This view has attractive features; it reduces the number of expected
but unobserved states. It emphasizes the view that quark-model
resonances, dynamically generated resonances, five-quark states are
different approaches to understand the same object with its
complicated internal structure. The different approaches are all
legitimate, none of them carries the truth, and each approach should
be tested if the predictions are not in conflict with firm results
of other approaches. If the view is extended to the mesonic sector,
there are some unfamiliar conclusions. First, the raison d'etre of
all scalar mesons is their $q\bar q$ component, even of the
$\sigma$. But this is also a trivial statement: without QCD, there
are no nuclear forces. Second, if there is only one scalar isoscalar
state (plus radial excitations), with all configurations $q\bar q$,
$qq\bar q\bar q$, $q\bar qG$, $\pi\pi$, $K\bar K$, etc. included,
then one further possibility, the $GG$ glueball, is likely also a
Fock component in its wave function. In this view, also the $GG$
component does not lead to an extra state, there is no
supernumerocity of resonances expected. I am aware of claims that
supernumerocity has been proven; the Particle Data Group is strongly
biassed into this direction. I maintain that these claims are
experimentally not sound, see [20], for a review. Presumably, most
scalar isoscalar mesons are realized as flavor singlets or octets.
The SU(3) flavor singlets are very wide and form the continuous
scalar background, a ``narrow" $f_0(1370)$ does not exist. Only the
octet mesons have normal hadronic widths; these are $f_0(1500)$,
$f_0(1760)$ (which includes $f_0(1710)$, $f_0(1790)$, $f_0(1810)$),
and $f_0(2100)$. The flavor decomposition of the low-mass $f_0(980)$
and $f_0(500)$ are strongly influenced by the $\pi\pi$ and $K\bar K$
threshold. Scalar mesons have a significant four-quark component but
their $q\bar q$ component is mandatory: in SU(3), there are nine
$q\bar q$ states and also nine $qq\bar q\bar q$ states respecting
the Pauli principle. In SU(4), there are 16 $q\bar q$ and there
could exist 36 $qq\bar q\bar q$ states, but none of these 20
additional states has been observed.

Interestingly, the view forbids non-exotic hybrids, since they are
absorbed into the wave function of $q\bar q$ mesons carrying the
same quantum numbers. Neither the existence of exotic hybrids is
excluded by the arguments given above, nor the existence of
pentaquarks in a non-$(qqq)$ configuration.

\section{What is needed?}
{\bf It ain't necessarily be so: \\[0.5ex]}
The interpretation depends, of course, heavily on the existing data.
Hence it is of greatest importance to confirm or refute as many of
the states seen in KH and CM and not seen in GWU analyses as
possible. Photoproduction experiments start to have a significant
impact. Experiments with polarized photon beams and polarized
targets have already taken a significant amount of data; results are
eagerly waited for. Hyperon photoproduction experiments benefit from
the self-analysing power of hyperon decays. Some double polarization
variables hit the value 1. This may indicate that a smaller number
of observables (and not 8) is already sufficient to constrain the
amplitudes fully. At least, the BnGa PWA group noticed that there
are much less ambiguities in defining the partial wave amplitudes
for hyperon photoproduction when the amplitudes are constrained by
$d\sigma/d\Omega$, $\Sigma$, $P$, $C_x, C_z$, $O_x$, and $O_z$.
Hence I believe we are at a point where we soon will be able to
decide which resonances exist in the mass range below 2\,GeV or,
perhaps, 2.2\,GeV. \\[1ex]
\noindent{\bf Search for missing quark model states: \\[0.5ex]}
 In the second
oscillator shell with $L=2$, quark models predict states in which
the spatial wave function are fully antisymmetric, in which the
orbital angular momenta $\vec l_{\rho}$ and $\vec l_{\lambda}$ are
both one and couple to $L=1$. If these states exist, they likely do
not decay into $N\pi$ but prefer a cascade where the two oscillators
de-excite successively. According to the AdS/QCD mass formula (and
its interpretation) one can assume that the mass of this spin
doublet $N_{1/2^+}$ and $N_{3/2^+}$ should be between 1.7 and
1.8\,GeV. The preferred decay mode could be the cascade
$N_{1/2^+}\to N_{1/2^-}(1535)\pi\to N\eta\pi$ or $N_{3/2^+}\to
N_{3/2^-}(1520)\pi\to N\pi\pi$ where all decays proceed via S-wave. \\[1ex]
\noindent{\bf What is the mass of the first $\Delta_{7/2^-}$
resonance? \\[0.5ex]}
We have already mentioned the quartet of states
$\Delta_{1/2^+}(1910)$, $\Delta_{3/2^+}(1920)$, $\Delta
_{5/2^+}(1905)$, and $\Delta _{7/2^+}(1950)$. The first three states
have spin-parity partners $\Delta_{1/2^-}(1900)$,
$\Delta_{3/2^-}(1940)$, $\Delta _{5/2^-}(1930)$. In quark models,
the masses of positive parity states are well reproduced, those of
the negative-parity states are unexpectedly low. In AdS/QCD, the
positive-parity states have $L=2,N=0$, the negative-parity states
$L=1,N=1$, respectively, and are predicted to be degenerate in mass.
Since $\vec L+\vec S$ yields only $J=1/2^-$, $J=3/2^-$, and
$J=5/2^-$ negative-parity states, but $J=1/2^+$, $\cdots$ $J=7/2^+$
for $L=2,S=3/2$, the absence of a parity partner of $\Delta
_{7/2^+}(1950)$ is expected. Instead, 2184\,MeV is predicted as
$\Delta _{7/2^-}$ mass, close to the PDG resonance $\Delta
_{7/2^-}(2200)$.
 \\
If chiral symmetry were restored in high-mass baryons, the $\Delta
_{7/2^-}$ mass would need to be degenerate in mass with $\Delta
_{7/2^+}(1950)$. The mass is 2200\,MeV, hence we could conclude that
AdS/QCD is favored. \hspace{-3mm}\begin{minipage}[c]{0.62\textwidth}
However, the $\Delta _{7/2^-}(2200)$ mass determination is not very
reliable. In Table 4 we list the Particle Data group entries for
$\Delta _{7/2^-}(2200)$. The values a very consistent but the
resonance is given one-star only. In the GWU analysis [4], the state
is not seen. This is certainly not the level of confidence we need
in order to settle the question if chiral symmetry is restored or
broken in high-mass hadron resonances.

Spin-parity doublets are also observed in the meson spectrum, for
many mesons but not for those on the leading Regge trajectory. To
give an
\end{minipage}\hspace{3mm}
\begin{minipage}[c]{0.35\textwidth}
{\bf Table 4:} {\it PDG entries for $\Delta _{7/2^-}(2200)$.} \\
\begin{tabular}{ccc}
\hline\hline
Mass [MeV]&Width[MeV]&  Ref.\\
2200$\pm$80& 450$\pm$100 &[2]\\
2215$\pm$60& 400$\pm$100 & [3]        \\
2280$\pm$80& 400$\pm$150 & [21]\\\hline\hline
\end{tabular}
\end{minipage}
\noindent example: there is a $a_4(2040)$ meson with $J^{PC}=4^{++}$
but the lowest mass partner with $J^{PC}=4^{-+}$ is $\pi_4(2250)$,
and the same observation can be made for the lowest-mass states in
the series $J^{PC}=1^{--}$, $J^{PC}=2^{++}$,
$J^{PC}=3^{--}$,~$\cdots$. Hence spin-parity partners are missing
for the most important states. The following argument by Glozman
[22] suggests a dynamical reason why the postulated $J^{PC}=4^{-+}$
state at 2040\,MeV was not observed: mesons in the high-mass region
stem mostly from the QMC-St-Petersburg analysis of Crystal-Barrel
LEAR data on $\bar pp$ annihilation in flight. And the $\bar pp$
system couples to $J^{PC}=4^{++}$ with $L=4$ while for formation of
a $J^{PC}=4^{-+}$ state, $L=5$ is required. Hence $\pi_4(2250)$
could be suppressed. It is worthwhile to note that the observed
pattern is is expected in AdS/QCD. Mesons -- except scalar and
pseudoscalar mesons -- are well described by the formula in the
caption of Fig. 6, if the ''3/2" is replaced by ''1/2".

In photoproduction or pion-induced reactions, there is no such
angular momentum  suppression; hence the search for the lowest-mass
$\Delta _{7/2^-}$ resonance seems to be the most rewarding case to
decide if quark models or AdS/QCD describe best the mass spectrum,
or if chiral symmetry is restored at high excitation energies.\\[1ex]
\noindent{\bf Limits of dynamical generated resonances: \\[0.5ex]}
The range of applicability of the method to generate resonances
dynamically has to be understood. As experimentalist, I am ready to
believe that $N_{1/2^-}(1535)$ can be generated dynamically from its
decay products, also that it could have, at large distances, a
sizable five-quark or molecular component in its Fock-space
expansion. But it is difficult to accept that $N_{1/2^-}(1535)$ and
$N_{3/2^-}(1520)$ are fundamentally different objects. Oset in his
contribution to this conference [23] constructed scalar and tensor
mesons from vector-vector interactions. The isoscalar scalar mesons
are found at 1512\,MeV and 1726\,MeV. This looks like a success as
well as the tensor meson found at 1275\,MeV. However, the next
tensor at 1525 couples with similar strength to $K^*\overline K^*$,
$\omega\omega$, $\phi\omega$, $\phi\phi$, but not to $\rho\rho$.
This is in striking conflict with what we know about $f_2(1525)$: it
is a $s\bar s$ state, the tensor mesons have a mixing angle close to
the ideal one. OZI rule violating couplings like the one into
$\phi\omega$ are highly suppressed. Also the isovector states are a
failure. If $\rho\rho$ generates $f_2(1270)$, $\rho\omega$ must
generate $a_2(1320)$; the predicted lowest isoscalar tensor mass is
1567\,MeV and its scalar companion is predicted to have 1777\,MeV.
In my view, it is important to find out the conditions for a
meaningful unitarization of chiral amplitudes, and not to enjoy the
achievements and to neglect the failures.\\[1.ex]
\noindent{\bf Excited states on the lattice: \\[0.5ex]}
Lattice gauge calculation have entered the difficult task to explore
the spectrum of baryon resonances and to address their finite width.
In the Introduction, I showed how the mass of a quark evolves with
decreasing momentum transfer. The calculation was done for a quark
propagator irrespective of its environment. According to the
discussion above, the constituent quark mass should depend on its
neighborhood; it should be lighter in the nucleon than in the
$\Delta(1232)$, and massive in highly excited states. \\[-2.5ex]

\section{Conclusions}
Baryons have played a very important role in the development of
particle physics, starting from the - at that time - mysterious
proton and neutron magnetic moments, the discovery of the
$\Delta(1232)$ by Fermi and his collaborators, to the discovery of
SU(3) and the insight that quarks need an extra degree of freedom
which is now known as color. The concept of chiral symmetry and
chiral symmetry breaking is at the root of modern strong interaction
physics. With the new tools, in experiment and theory, we have the
chance for a new understanding of strong interaction dynamics in the
confinement region.\\[1.ex]
{\it I would like to thank B. Metsch, E. Oset, H. Petry, and J.M.
Richard for clarifying discussions, and all members of SFB/TR16 for
continuous encouragement. Financial support within the SFB/TR16 is
kindly acknowledged.}\vspace{-1.mm}
\section{References}
\small
\begin{enumerate}
\vspace{-1.mm}\item
  C.~Amsler {\it et al.}  [Particle Data Group],
  Phys.\ Lett.\  B {\bf 667}, 1 (2008).
\vspace{-1.mm}\item
  G.~H\"ohler, F.~Kaiser, R.~Koch and E.~Pietarinen,
  ``Handbook of Pion Nucleon Scattering,''
Physics Data, No.12-1 (1979). \vspace{-1.mm}\item
  R.~E.~Cutkosky {\it et al.},
  ``$\pi N$ Partial Wave Analysis,''
4th Int. Conf. on Baryon Resonances, Toronto, Canada, 1980.
\vspace{-1.mm}\item
  R.~A.~Arndt, W.~J.~Briscoe, I.~I.~Strakovsky and R.~L.~Workman,
  Phys.\ Rev.\  C {\bf 74}, 045205 (2006).
\vspace{-1.mm}\item
  P.~O.~Bowman {\it et al.}, 
  Phys.\ Rev.\  D {\bf 71}, 054507 (2005).
  \vspace{-1.mm}\item
  A.~Chodos, R.~L.~Jaffe, K.~Johnson and C.~B.~Thorn,
  Phys.\ Rev.\  D {\bf 10}, 2599 (1974).
\vspace{-1.mm}\item
  M.~S.~Bhagwat, M.~A.~Pichowsky, C.~D.~Roberts and P.~C.~Tandy,
  Phys.\ Rev.\  C {\bf 68}, 015203 (2003).
\vspace{-1.mm}\item
  D.~Diakonov and V.~Y.~Petrov,
  Sov.\ Phys.\ JETP {\bf 62}, 204 (1985)
  [Zh.\ Eksp.\ Teor.\ Fiz.\  {\bf 89}, 361 (1985)].
\vspace{-1.mm}\item
  E.~Klempt and J.~M.~Richard,
  arXiv:0901.2055 [hep-ph].
\vspace{-1.mm}\item
  M.~D\"oring and K.~Nakayama,
  arXiv:0906.2949 [nucl-th].
\vspace{-1.mm}\item
  L.~Y.~Glozman,
  Phys.\ Rept.\  {\bf 444}, 1 (2007).
\vspace{-1.mm}\item
  L.~Y.~Glozman,
  arXiv:0903.3923 [hep-ph].
\vspace{-1.mm}\item
  G.~F.~de Teramond and S.~J.~Brodsky,
  arXiv:0903.4922 [hep-ph].
\vspace{-1.mm}\item
  S.~Capstick and N.~Isgur,
  Phys.\ Rev.\  D {\bf 34}, 2809 (1986).
\vspace{-1.mm}\item
  U.~L\"oring, B.~C.~Metsch and H.~R.~Petry,
  Eur.\ Phys.\ J.\  A {\bf 10}, 395 (2001).
\vspace{-1.mm}\item
  M.~P.~Mattis and M.~Karliner,
  Phys.\ Rev.\  D {\bf 31}, 2833 (1985).
\vspace{-1.mm}\item
  H.~Forkel and E.~Klempt,
  Phys.\ Lett.\  B {\bf 679}, 77 (2009).
\vspace{-1.mm}\item E. Klempt, Nucleon excitations, QNP2009,
arXiv:1001.2997v1 [hep-ph]. \vspace{-1.mm}\item B.S. Zou, ''Evidence
for a new $\Sigma^*$ resonance with $J^P=1/2^-$ around 1380 MeV ",
contr. 62. \vspace{-1.mm}\item
  E.~Klempt and A.~Zaitsev,
  Phys.\ Rept.\  {\bf 454}, 1 (2007).
\vspace{-1.mm}\item
  A.~W.~Hendry,
  Phys.\ Rev.\ Lett.\  {\bf 41}, 222 (1978).
\vspace{-1.mm}\item
  L.~Y.~Glozman,
  arXiv:0912.1129 [hep-ph].
\vspace{-1.mm}\item E. Oset, ''The $f_0(1370)$, $f_0(1710)$,
$f_2(1270)$, $f_2'(1525)$ and $K_2^*(1430)$ in a unitary approach",
contr. 51.
\end{enumerate}

\end{document}